\documentclass{optica-article}

\journal{opticajournal} 

\articletype{Research Article}

\usepackage{lineno}

\begin{document}

\title{High-precision measurements of terahertz polarization states with a fiber coupled time-domain THz spectrometer}

\author{Zhenisbek Tagay,\authormark{1} Ralph Romero III,\authormark{1} and N.~P.~Armitage\authormark{1,*}}

\address{\authormark{1}Department of Physics and Astronomy, The Johns Hopkins University, Baltimore, MD 21218 USA}

\email{\authormark{*}npa@jhu.edu} 


\begin{abstract*} 
We present a new method for high precision measurements of polarization rotation in the frequency range from 0.2 to 2.2 THz using a fiber coupled time-domain THz spectrometer.  A free standing wire-grid polarizer splits THz light into orthogonal components which are then measured by two separate detectors simultaneously. We theoretically model the uncertainties introduced by optical component non-idealities and predict that we may expect to achieve precisions of order 2.3 $\mu$rad (0.13 mdeg) and accuracies of 0.8$\%$ when anti-symmetrizing the response with respect to an applied field.  Anti-symmetrization improves precision by more than four orders of magnitude.  We demonstrate this method on a 2D electron gas in magnetic field and show that we achieve a precision of 20 $\mu$rad (1.1 mdeg) for small polarization rotation angles. A detailed description of the technique and data analysis procedure is provided, demonstrating its capability to precisely measure polarization states in the 0.2 to 2.2 THz range.

\end{abstract*}

\section{Introduction}
Highly accurate polarization state measurements in THz range are of fundamental interest for condensed matter physics as material systems may exhibit very small Kerr or Faraday rotations that are diagnostic of their material properties~\cite{shimano_graphenee, Kapitulnik_sro,armitage2014constraints, BiNi, shimano_sro, wu_bise, chaudhuri_bst}. Despite extensive efforts, polarization-dependent measurements in this frequency range have proven be difficult due to the many instrumentation challenges~\cite{castro}. This includes insufficient signal-to-noise ratios (SNR) of THz light sources, finite extinction ratios of wire-grid polarizers (WGP), the overall quality of THz lenses and mirrors, and the inability to properly align these optical components.  One of the earlier approaches to accurately measure the polarization rotation was to use two WGPs in crossed-Nicol configuration~\cite{Kanda_crossNicol, Nagashima_crossNicol, shimano_graphenee}. The ultimate precision of this techniques was reported to be 0.5 mrad (29 mdeg). Despite its advantages, this method involves 90$^\circ$ rotation of WGP between orthogonal polarization measurements and, therefore, its sensitivity is limited by the extinction ratio of the WGP and the angular precision of the rotation stages used in the setup.  Our group  previously developed a polarization modulation technique to simultaneously measure both orthogonal components of THz pulse~\cite{morris_modulation}. We had used a fast-rotating polarizer to modulate the signal and measured it with lock-in technique. This method allows one to measure small polarization rotations with an accuracy of 0.9 mrad (52 mdeg) and a precision of 0.35 mrad (20 mdeg). Although this technique eliminates the effects of finite extinction ratio of WGPs discussed above, its sensitivity is instead limited by mechanical robustness of polarizers to high-speed rotations. Other work has used electro-optic sampling as a detection scheme and modulated the probe-polarization to measure small rotations~\cite{nemoto_probemodulation}. Due to the high SNR of intense THz light generated by the optical rectification method the resulting precision of the polarization state measurements was significantly improved and was reported to be 0.1 mrad (6 mdeg). However, the reported value corresponds to a sensitivity as measured at the peak of THz signal where SNR is the highest, not the entire pulse. Therefore, the overall precision is likely not as good overall as reported.

Recent developments in THz spectroscopy using fiber coupled laser and InGaAs photo-conductive antennas have demonstrated significantly improved SNR and much faster data acquisition times~\cite{THz_antenna,toptica_90dBB}. These unprecedented capabilities present opportunities to study optical properties of materials with much higher precision than before.  With this in mind we have built a high-precision time-domain THz spectroscopy (TDTS) and polarimetry setup using a fiber-coupled laser-based spectrometer. Commercially available WGPs with the best available extinction ratios were utilized to accurately control the polarization state of light. The THz signal transmitted through the sample is split into two orthogonal components with a WGP and measured by two separate detectors simultaneously. An extensive calibration scheme is performed to account for non-idealities in the detection channels. Nevertheless, due to high SNR and absence of moving optical parts we ultimately achieve the relative accuracy of 0.8\% and precision of 0.02 mrad (1.1 mdeg) in small rotation angle measurements in the range of 0.3-1 THz. This level of sensitivity provides an excellent opportunity to investigate entirely new classes of materials and phenomena that were not accessible with previous techniques. In this work, we describe the optical setup, data acquisition procedure, and discuss some initial measurements performed with it. 

\section{Instrumentation} 
\begin{figure}[b!]
\centering
\includegraphics[width=\textwidth]{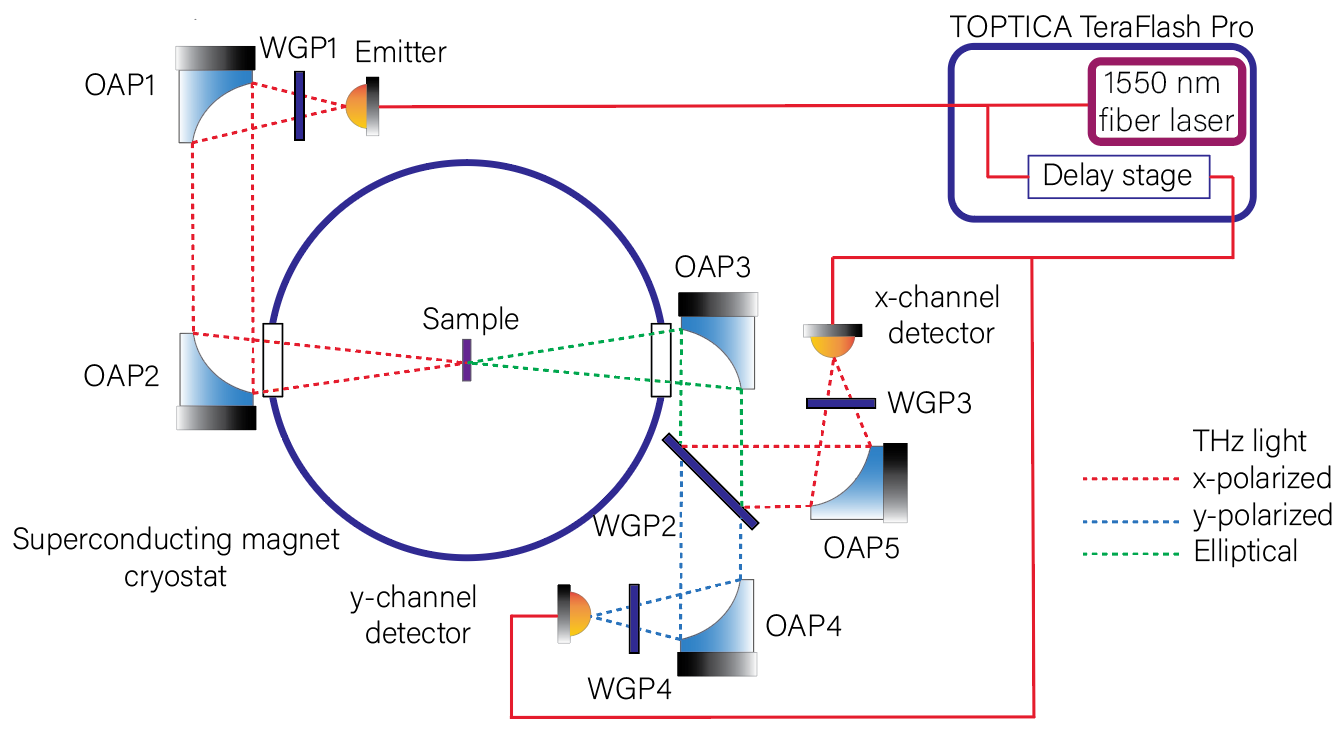} 
\label{Fig.Conductivity}
\caption{Experimental setup}
\end{figure}
The experimental setup is shown in Fig.~1. The generation and detection of THz pulses is done by a Toptica TeraFlash Pro spectrometer. It couples a 1550 nm fiber laser via a fiber optic coupling to InGaAs based photo-conductive switches to produce THz radiation with a dynamic range of 95 dB (50,000:1 SNR).  The spectrometer is capable of 100 dB peak dynamic range and greater than 6 THz bandwidth.  A ``fast shaker" mechanical delay stage allows 60 traces/sec to be taken for typical 15 ps scans. The maximum scan length supported by the delay stage is 200 ps.  Our particular spectrometer was commissioned with one emitter and two detector switches that are activated by the same laser.  We mount the emitter switch onto a rotation stage, which allows the primary polarization to be adjusted.  We typically align it horizontally.  WGP1 with a horizontal transmission axis is placed right in front of it to further and accurately control the polarization state. WGPs used in the setup were obtained from PureWavePolarizers Ltd and have extinction ratio of $10^{-2}$ at 1 THz. They were mounted onto rotation stages and all aligned to the same angular reference using interference patterns from shining a HeNe red laser on to them. To reduce errors, the alignment was done after the WGPs were mounted onto their final positions in the optical setup. The off-axis parabolic mirrors (OAPs) are laid out in an 8$f$-configuration~\cite{8fgeom}. OAPs 1 and 2 are used to collimate and focus the THz beam. The horizontally polarized light passes through the sample, where it -- generally -- becomes elliptically polarized according to the sample's transfer (${\hat T}$) matrix. The resulting elliptically polarized light is then collimated by OAP3.  WGP2 with vertical axis of transmission is oriented at 45$^{\circ}$ with respect to the incident beam propagation direction to split the THz pulse into two orthogonal components. The vertical component of THz light passes through WGP2, gets focused by OAP4  and is detected by the photo-conductive switch. The horizontal component is reflected by  WGP2 in the perpendicular direction, focused by OAP5 and detected by another photo-conductive switch. The detectors measure in a polarization resolved manner, but WGP3 and WGP4 with horizontal and vertical transmission axes, respectively, are placed in front of the detectors and used to filter out any leaked signal from WGP2.   Samples were mounted and held inside a closed cycle Janis cryostat with two sets of internal windows and a superconducting magnet.

\section{Data analysis and corrections }
\subsection{Calibration of detectors}
As mentioned in the instrumentation section, two separate THz detectors are used to measure horizontal ($\hat{x}$) and vertical ($\hat{y}$) components of the THz light. Although these detectors are nominally the same, their sensitivities and alignments invariably will be slightly different.  Thus, calibration of channels has to be done for accurate polarization measurements.  We make use of the Jones matrix formalism~\cite{Jones} as a mathematical basis for the calibration scheme. Suppose the transmission through the entire optical system (including all lenses, mirrors and windows) is denoted by a matrix $\hat{M}$ and  sensitivities of $\hat{x}$- and $\hat{y}$-detection channels are $S_x$ and $S_y$, respectively. 

For an incident polarization of $\begin{bmatrix}E_x^0\\E_y^0\end{bmatrix}$, the corresponding readout on detectors would be:
\begin{equation}   \begin{bmatrix}E_x\\E_y\end{bmatrix}=\begin{bmatrix}S_{x} & 0\\0 & S_{y}\end{bmatrix}\big[\hat{M}\big]\begin{bmatrix}E_x^0 \\E_y^0 \end{bmatrix}
\end{equation}

In order to extract the sensitivity ratio of two detectors $S_x/S_y$, we do 2 sets of measurements with different incident polarizations: $45^{\circ}$ and $-45^{\circ}$ with respect to $\hat{x}$-axis. The corresponding readouts on the detectors are:
\begin{equation}
\begin{split}
    \begin{bmatrix}E_x^{45}\\E_y^{45}\end{bmatrix} &=\begin{bmatrix}S_{x} & 0\\0 & S_{y}\end{bmatrix}\big[\hat{M}\big]\begin{bmatrix}\frac{E_0(1+\eta)}{2} \\\frac{E_0(1-\eta)}{2} \end{bmatrix}
\\
    \begin{bmatrix}E_x^{-45}\\E_y^{-45}\end{bmatrix} &=\begin{bmatrix}S_{x} & 0\\0 & S_{y}\end{bmatrix}\big[\hat{M}\big]\begin{bmatrix}\frac{E_0(1+\eta)}{2} \\\frac{-E_0(1-\eta)}{2} \end{bmatrix}
\end{split}
\end{equation}
where $E_0$ is electric field amplitude of incident THz beam and $\eta$ is the extinction ratio of WGP1. Polarization states $\frac{E_0}{2}\begin{bmatrix}1+\eta\\1-\eta\end{bmatrix}$ and $\frac{E_0}{2}\begin{bmatrix}1+\eta\\-(1-\eta)\end{bmatrix}$ are carefully prepared by orienting the THz emitter parallel to $\hat{x}$-axis and using a wire-grid polarizer mounted on a motorized rotation stage in front of it to control the polarization. Transmission matrix for a WGP with finite $\eta$ will be given later in Eq.~\ref{polarizer}.  Under the assumption that transfer matrix $\hat{M}$ can be approximated as an identity matrix, the sensitivity ratio can be expressed as follows:
\begin{equation}
    \frac{S_x}{S_y} \approx\sqrt{\frac{|E_x^{45}E_x^{-45}|}{|E_y^{45}E_y^{-45}|}}\bigg[\frac{1-\eta}{1+\eta}\bigg]
    \label{magnitudecorrection}
\end{equation}
In the limit $\eta\rightarrow 0$, the final factor in Eq.~\ref{magnitudecorrection} converges to 1 and the the expression for $S_x/S_y$ becomes independent of $\eta$. Therefore, it is essential to use high-quality WGPs with low $\eta$ for accurate measurements of the sensitivity ratio.  Without good polarizers this calibration can be greatly effected.  Commercially available WGPs in THz range typically have extinction ratios on the order of $10^{-2}$. Neglecting the $\eta$-dependence in this case would lead to $\approx$ 2\% relative uncertainty in $S_x/S_y$. To account for the effect of finite extinction ratios we can perform an additional measurement with incident polarization of 0$^{\circ}$ with respect to $\hat{x}$-axis. After a few algebraic transformations, one can rewrite Eq.~\ref{magnitudecorrection} purely in terms of quantities from direct measurements:
\begin{equation}
    \frac{S_x}{S_y} \approx \sqrt{\frac{|E_x^{45}E_x^{-45}|}{|E_y^{45}E_y^{-45}|}}\bigg[\frac{2|E_x^0|-|E_x^{45}|-|E_x^{-45}|}{|E_x^{45}|+|E_x^{-45}|}\bigg]
    \label{magnitudecorrection_new}.
\end{equation}

\begin{figure}[b!]
\centering
\includegraphics[width=0.6\textwidth]{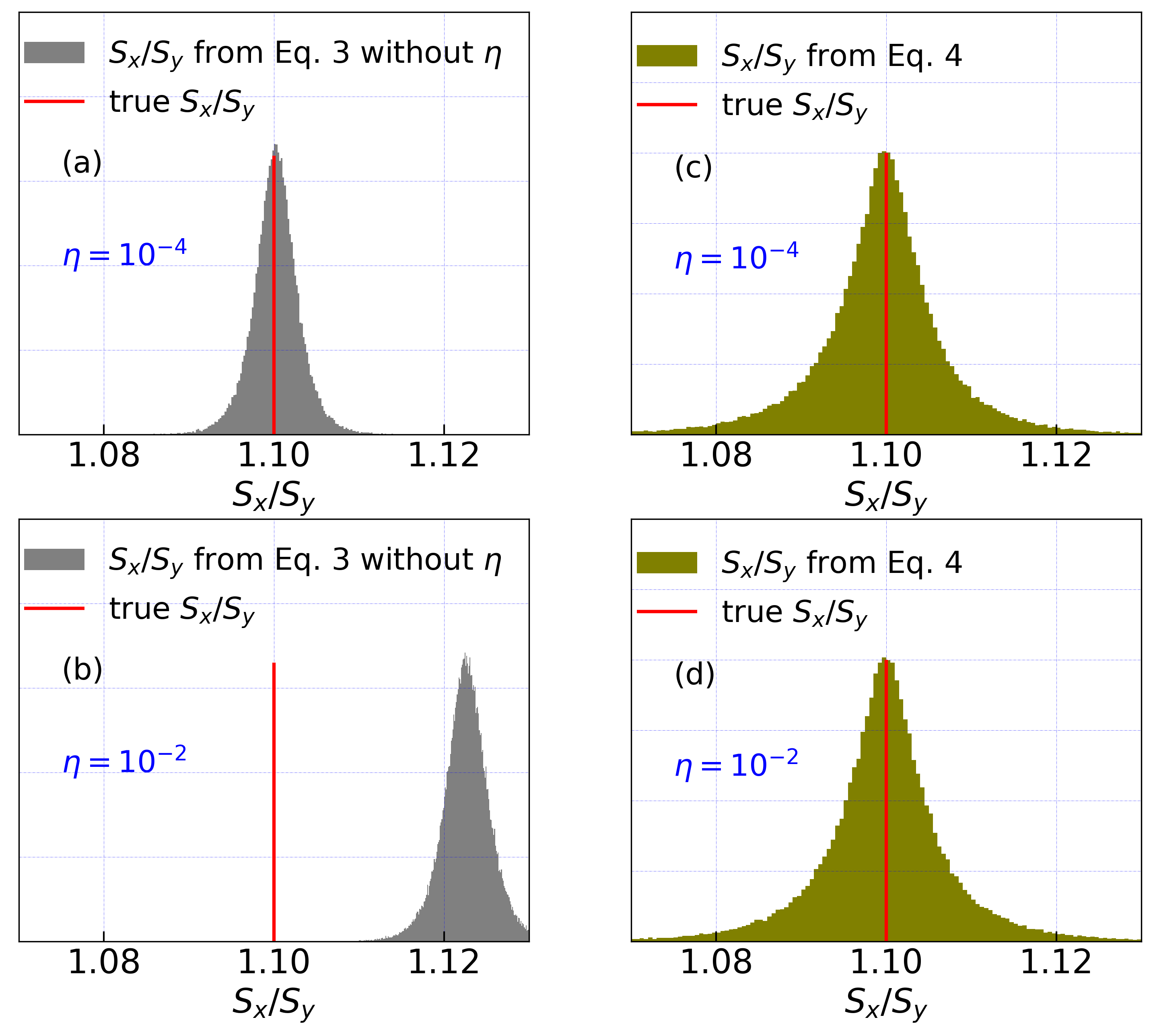} 
\label{Fig.calibration}
\caption{Numerical simulations of $S_x/S_y$: by neglecting the finite extinction ratio in Eq.~\ref{magnitudecorrection} for (a) $\eta=10^{-4}$ (b) $\eta=10^{-2}$, and (c)-(d) by using Eq.~\ref{magnitudecorrection_new} for the same values of $\eta$}.
\end{figure}
To validate the accuracy of Eq.~\ref{magnitudecorrection_new} under realistic conditions, we run numerical simulations with reasonable estimates for off-diagonal elements of matrix $\hat{M}$. The details of these estimations are provided in Section 4.  We ran 100,000 realizations and plot the results in Fig. 2.  Simulations are done for the case when sensitivities differ by 10\% (i.e. $S_x/S_y=1.1$).
In Figs. 2(a)-(b) we first plot the $S_x/S_y$ calculated by neglecting the finite extinction ratio effect in Eq.~\ref{magnitudecorrection}. Simulations for the case $\eta=10^{-4}$ result in gaussian-like distribution centered around the expected value of 1.1 with half-width of 0.003. However, the distribution for $\eta=10^{-2}$ is shifted from the true value by the significant amount of 0.02. This observation supports our earlier statement that ignoring $\eta$ in calculation of $S_x/S_y$ leads to significant deviations from the true value when $\eta$ is relatively large. In Figs. 2(c)-(d) we now show the sensitivity ratio calculated using Eq.~\ref{magnitudecorrection_new}. In both cases, simulations result in gaussian-like distribution of $S_x/S_y$ centered {\it around} the expected value of 1.1 regardless of $\eta$. The distributions for both $\eta=10^{-4}$ and $\eta=10^{-2}$ have a half-width of about 0.008, which correspond to the relative uncertainty of about 0.8\% in calculating the sensitivity ratio between two orthogonal channels. This uncertainty can be attributed to finite off-diagonal elements of matrix $\hat{M}$ due to non-ideal optical elements which will be addressed below. Nevertheless, numerical simulations indicate that Eq.~\ref{magnitudecorrection_new} can be used to accurately measure $S_x/S_y$ with relatively small error.

During analysis, the electric field  measured in the $\hat{y}$-channel has to be multiplied by the sensitivity ratio before further analysis to get accurate results. The calibration procedure described above should be performed only on frequency-domain signals (i.e. Fourier transforms of time traces), not time traces themselves. Using time-domain data for calibration is incorrect and most likely will result in algebraic errors (e.g. division by 0).   Also note in Eq.~\ref{magnitudecorrection} and Eq.~\ref{magnitudecorrection_new} above we have taken the absolute values of the signals such that the procedure only provides a correction for magnitudes.   We discuss the phase corrections below.

 \subsection{Correcting for phase shifts in data}
Since we are interested in complex response functions (not just their amplitudes),  phase related corrections must be made. First, there is a physical path length difference between the two detectors. This will result in a phase difference $\Delta\phi_{x,y}=\omega\Delta t_{x,y}$, where $\Delta t_{x,y}$ is the path difference in the time-domain between $\hat{x}$ and $\hat{y}$ channels. This effect can be reduced to a minimum by adjusting the relative positions of the two detectors, but it is limited by the precision of the mechanical translation stages. Therefore, post-processing of the data is needed to account for remaining path length difference.  In addition, it was observed that fiber-coupled THz spectrometers suffer from drift of the signal's measure phase in a fashion that does not occur for free space coupled systems.  The origin of these time shifts are unclear, but could be originating from finite positional accuracy of the fast-shaker  delay stage or acoustic vibrations being coupled into the optical fibers. The observed phase shifts are linear in frequency and result in the peak position moving in the time-domain. Since both systematic effects are of the same nature (i.e. have form $\Delta\phi=\omega\Delta t$), in principle they can be remedied in a similar fashion. 

To account for path difference between two channels we first set the incident polarization to $45^{\circ}$ such that both $x$ and $y$ signal have same amplitude and before starting measurements measure the time difference $\Delta t_{x,y}=t_x-t_y$ between their signal peaks. This is then used as a global time reference to calculate $\Delta \phi_{x,y}$. This phase difference can be measured on substrates or aperture references at the calibration stage and is used as an initial phase calibration for the rest of the measurements in the data set. 

\begin{align}
    E_x&=E_x^{raw} & E_y&=E_y^{raw}\exp(i\Delta\phi_{x,y})
\end{align}

We have found the second type of phase shift rather difficult to deal with since it varies with time in the TeraFlash system with about 10 fs variation on 30 sec time scales. To correct for time-dependent phase shifts one needs to continuously align both $\hat{x}$ and $\hat{y}$ signal positions to a global reference.  To do that, we measure signal peak positions though a reference (substrate/aperture) $t_x$ and $t_y$ with incident polarization set to $0^{\circ}$ and use the small leakage signal into $\hat{y}$ to get the relative time shift.  During the actual experiment we deal with these time dependent time shifts by measuring the sample and reference time traces back to back. Suppose at instance $i$ the peak positions of $\hat{x}$ and $\hat{y}$ signals as measured on a reference are $t_x^i$ and $t_y^i$. The corresponding phase shifts are given by $\Delta \phi_x^i=\omega (t_x-t_x^{i})$ and $\Delta \phi_y^i=\omega (t_y-t_y^{i})$. These values are then used to phase align the Fourier transforms of the data.

\subsection{Polarization rotations due to non-ideal optical elements and alignment}
The main source of uncertainty in these polarimetry measurements are systematic errors due to non-ideal optical components and their alignment. Lenses, mirrors and polarizers used in the THz range can be far less accurate than their counterparts in the IR range. Except for the random phase fluctuations discussed above, random errors are not particularly significant in the present case due to the high signal-to-noise ratio of the spectrometer.  A significant source of systematic errors can be finite rotations from small misalignments of the off-axis parabolic mirrors (OAPs) that are used to collimate/focus THz beams. Typical OAPs in THz experiments can rotate the polarization of a light by up to few degrees if misaligned. The Jones matrix for rotation from such an OAP is given by:

\begin{equation}
    {\hat R}(\phi)=\begin{bmatrix}\cos{(\phi)} & \sin{(\phi)}\\-\sin{(\phi)} & \cos{(\phi)}\end{bmatrix},
\end{equation}
where $\phi$ is the rotation angle.

Another source of systematic error originates from non-ideal wire-grid polarizers (WGPs) used in the setup. Typical WGPs in the THz range have a finite extinction ratio $\eta$.  This sets a significant limitation on overall performance of this spectrometer. In addition, small miscalibrations of the WGP angle can introduce additional systematic errors. The transmission matrix for a WGP with an extinction ratio of $\eta$ and at an angle $\theta$  with respect to horizontal axis is represented by:

\begin{equation}
{\hat P}(\eta,\theta)=\begin{bmatrix}\cos^2{(\theta)}+\eta\sin^2{(\theta)} & (1-\eta)\cos{(\theta)}\sin{(\theta)}\\(1-\eta)\cos{(\theta)}\sin{(\theta)} & \sin^2{(\theta)}+\eta\cos^2{(\theta)}\end{bmatrix}
\label{polarizer}
\end{equation}

Due to non-idealities of the optical components or their alignments, one observes a finite vertical field even if the emitted THz beam is strictly horizontal and the sample itself does not rotate light.  This means that during actual experiments, real Faraday rotation from a sample will mix with background rotation through the optical system making it challenging to obtain accurate measurements.  Ideally one would be able to ``reference out" these non-idealities of the spectrometer as one does in conventional TDTS where rotations are small and can be neglected.  In conventional TDTS one computes the transmission function by dividing transmission through the sample by transmission through a bare substrate (for thin films), or a blank aperture (for single crystals).   However this is not generally possible for cases where even very small rotations are to be measured.  The issue can be seen as follows.  The most general expression for transmission through the entire spectrometer is

\begin{equation}
\begin{bmatrix}E_x\\E_y\end{bmatrix}= 
\begin{bmatrix}A_{xx} & A_{xy}\\A_{yx} & A_{yy}\end{bmatrix}
\begin{bmatrix}T_{xx} & T_{xy}\\T_{yx} & T_{yy}\end{bmatrix}  
\begin{bmatrix}B_{xx} & B_{xy}\\B_{yx} & B_{yy}\end{bmatrix}
\begin{bmatrix}E^0_x \\E^0_y \end{bmatrix},
\end{equation}
where $\vec{E}_0$ is the incoming field, the matrix $\hat{B}$ represents all rotations and projections before the sample, $\hat{T}$ is the sample Jones matrix, and $\hat{A}$ represents all rotations and projections after the sample. $\vec{E}$ is the measured THz field.  One can also define a ``reference field" in the same spirit as for normal TDTS as the transmission through the system, but without the sample.  It is:
\begin{equation}
\begin{bmatrix}R_x\\R_y\end{bmatrix}= 
\begin{bmatrix}A_{xx} & A_{xy}\\A_{yx} & A_{yy}\end{bmatrix}
\begin{bmatrix}B_{xx} & B_{xy}\\B_{yx} & B_{yy}\end{bmatrix} 
\begin{bmatrix}E^0_x \\E^0_y \end{bmatrix}.
\end{equation}

Unfortunately a relation in the spirit of conventional THz does not apply e.g. the below is an {\it inequality}

\begin{equation}
\begin{bmatrix}E_x\\E_y\end{bmatrix} \neq  \begin{bmatrix}T_{xx} & T_{xy}\\T_{yx} & T_{yy}\end{bmatrix}  \begin{bmatrix}R_x \\R_y \end{bmatrix}.
\label{wrongmatrix}
\end{equation}
The problem is mathematical and arises because of the general non-commutativity of matrices.  Matrices only commute if they are simultaneously diagonalizable e.g. a common rotation for all elements would render them diagonal.   That is certainly not the case here as optical elements can be misaligned independently.  Nevertheless we believe a related form of Eq. \ref{wrongmatrix} is {\it approximately} true when applied to data that is anti-symmetrized with respect to the sign of the magnetic field (or the sense of the spontaneous time-symmetry breaking).   Reversing the direction of the magnetic field is equivalent to reversing the direction of time.  Using the relations of Ref.~\cite{armitage2014constraints} for the time-reversed transmission matrix of a system with $C_4$ symmetry, the off-diagonal elements change sign upon time-reversal.  We can calculate

\begin{equation}
\vec{E} \approx  \frac{1}{2}   [ \hat{T}(B) -  \hat{T}(-B)   ]   \vec{R} .
\label{approxmatrix1}
\end{equation}

\begin{figure}[b!]
\centering
\includegraphics[width=0.6\textwidth]{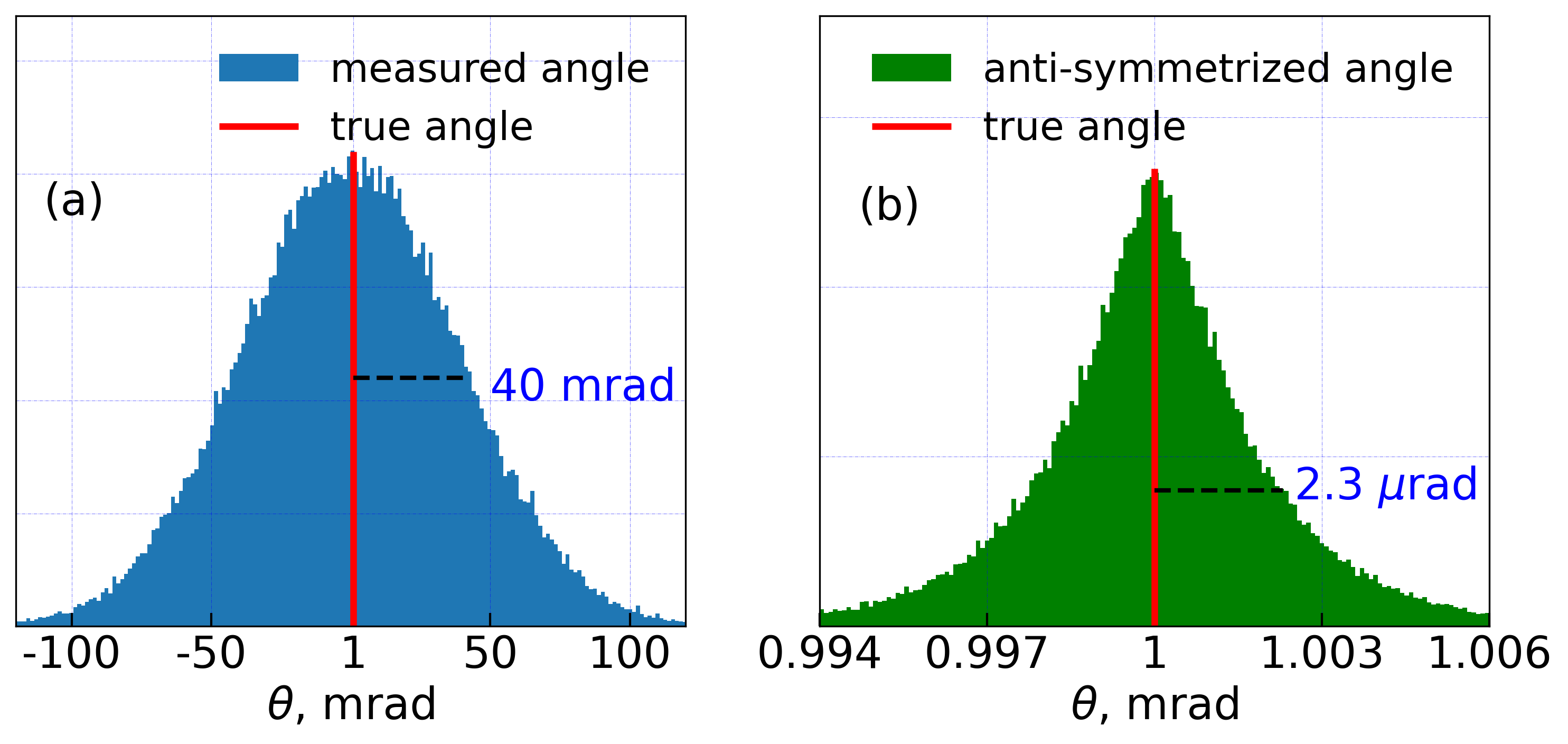} 
\label{Fig.Conductivity}
\caption{Simulation of Faraday rotation in non-ideal setup using a distribution of misalignments specified in the text: (a) measured Faraday rotation, and (b) anti-symmetrized Faraday rotation. Red line indicates the expected Faraday rotation from the sample.}
\end{figure}

To use this ``anti-symmetrized" expression, one measures $\hat{x}$ and $\hat{y}$ electric fields for the sample and reference and corrects the amplitude and phase using the above procedures.   This gives $E_x$, $E_y$, $R_x$ and $R_y$.  In materials preserving $C_4$-symmetry, diagonal elements of $\hat{T}$ are equal ($T_{xx}=T_{yy}$) and off-diagonal elements have same magnitude but opposite sign ($T_{xy}=-T_{yx}$)~\cite{armitage2014constraints}.   Then given Eq. \ref{approxmatrix1} one can derive expressions for the transmission matrix elements:
\begin{align}
    T_{xx}& \approx \frac{E_x R_x +E_yR_y}{R_x^2+R_y^2} & T_{yx}& \approx \frac{E_y R_x -E_xR_y}{R_x^2+R_y^2}
    \label{approxmatrix2}
\end{align}
Faraday rotation through a sample can be expressed as $\theta=T_{yx}/T_{xx}$.

\section{Simulations}

As long as misalignments are small, the errors incurred in Eqs. \ref{approxmatrix1} and \ref{approxmatrix2} when performing the anti-symmetrization procedure are very small.   Unfortunately, it is fairly involved to show 
analytically how the anti-symmetrization procedure largely gets rid of systematic backgrounds since it involves multiplication of several Jones matrices with finite off-diagonal elements. Instead, we can run numerical simulations with reasonable estimates for the non-idealities described above. Suppose we are studying a sample which should hypothetically give a 1 mrad Faraday rotation (e.g. $\theta=T_{xy}/T_{xx}=0.001$) and which is odd in applied field.  We do a simulation, where we allow polarization rotations from the four OAPs to take values up to $1^{\circ}$, the four WGPs to have angular misalignments up to $1^{\circ}$, and the emitter, and detectors to be $5^{\circ}$ misaligned.  We did not include the effects of cryostat windows, but they are expected to introduce errors much the same way as the other components.  These represent the rough estimates of the errors in alignment, which were limited by the polarizer rotation stages and mirror mounts used for the detectors.   The simulation used approximately 100,000 realizations with alignments taken from Gaussian distributions with the above given half-widths for the misalignments.  The total rotation of the beam due to the accumulated rotations for particular realization of misalignments is generally of order 40 mrad as can be seen in Fig. 3(a).  However, anti-symmetrization reduces this error by more than four orders of magnitude.  As shown in Fig. 3(b), the numerical simulation for anti-symmetrized Faraday rotation is a gaussian-like distribution with the expected mean of 1 mrad and half-width of 2.3 $\mu$rad (0.13 mdeg).  This error is well below the random noise floor of the spectrometer, therefore the anti-symmetrization procedure effectively reduces the systematic uncertainty and enables us to accurately measure the polarization state of THz light.

\section{Applications}
To test the capabilities of the spectrometer we measure the Faraday rotation though a InGaAs 2D-electron gas system in applied magnetic field. This sample was previously studied in Ref.~\cite{chauhan_2deg} using the polarization modulation technique~\cite{morris_modulation}. In Fig.  4(a)-(b) we present real and imaginary parts of the Faraday rotation up to a field of 3 T at T = 6K. The shape and field-dependence of the rotation is in good agreement with cyclotron resonance behavior described in the original work. Performing the same analysis we find that the electron gas has a carrier density of $n=2.5\times 10^{11}$/cm$^2$ with effective mass $m=0.039$ $m_e$ and scattering rate of $\Gamma/2\pi=0.55$ THz.  In Fig. 4(c) we plot the standard deviation of the complex Faraday rotations from three subsequent measurements measured back to back to show the precision of the experimental data. Uncertainty for the measured angle between 0.3-1 THz was measured to be 0.02 mrad (1.1 mdeg), and doesn't exceed the value of 0.1 mrad for the entire spectral range of 0.2-2.2 THz.   This data was anti-symmetrized according the procedures discussed above.

\begin{figure}[t!]
\centering
\includegraphics[width=\textwidth]{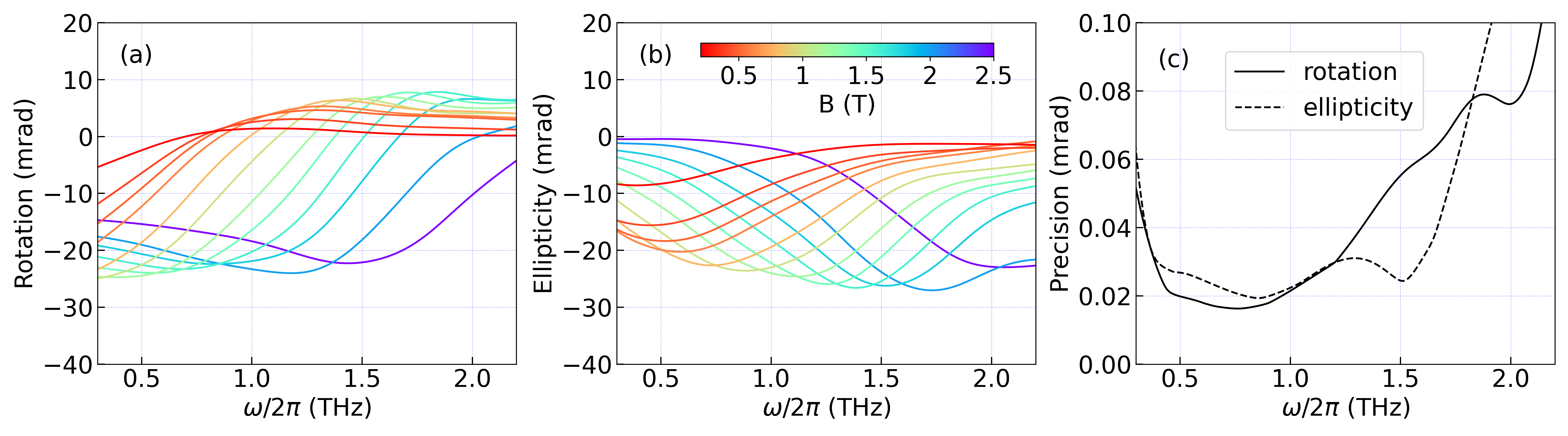} 
\label{Fig:Misalign}
\caption{Real (a) and (b) imaginary part of Faraday rotation in InGaAs. (c) Standard deviation of measured Faraday rotation.}
\end{figure}

\section{Conclusions}
In this study, we have developed a high-precision time-domain THz polarimetry setup using fiber-coupled spectrometer. Due to simultaneous detection of both orthogonal components of THz beam and fiber coupling we achieved an excellent relative accuracy of 0.8\% and precision of 0.02 mrad in small angle measurements. This unprecedented sensitivity will provide opportunity to study entirely new classes of materials and can be employed in various applications.

\section*{Acknowledgements}
This work was supported by the ARO ``MURI: Implementation of axion electrodynamics in topological films and devices" and the Gordon and Betty Moore Foundation EPiQS Initiative Grant GBMF-9454 to NPA

\bibliography{main}
\end{document}